\magnification=\magstep1
\hfuzz=6pt
\baselineskip=16pt
$ $

\vskip 2cm

\centerline{\bf Universal quantum computation in integrable systems}

\bigskip
\centerline{Seth Lloyd$^1$ and Simone Montangero$^2$}

\bigskip
\centerline{1. Department of Mechanical Engineering,
Research Laboratory for Electronics}

\centerline{ Massachusetts Institute of Technology, 
3-160, Cambridge MA 02139 USA}

\centerline{2. Institut f\"ur Quanteninformationsverarbeitung
Universit\"at Ulm \& IQST, 89069 Ulm, Germany}

\bigskip\noindent{\it Abstract:} 
Quantized integrable systems can be made to perform
universal quantum computation by the application of a global
time-varying control.  The action-angle variables of the integrable
system function as qubits or qudits, which can be coupled selectively
by the global control to induce universal quantum logic gates.  By contrast,
chaotic quantum systems, even if controllable, do not generically allow
quantum computation under global control.

\vskip 1cm

In classical mechanics, integrable systems are ones that are
dynamically `well-behaved.'  Their dynamics are characterized
by a set of conserved quantities, and are stable under small 
perturbations [1].  The quantized versions of integrable linear systems
inherit much of the good behavior of their classical counterparts [2-5].
In this paper, we investigate the problem of controlling quantized
integrable systems and show that generically, quantized integrable
systems can be made to perform universal quantum computation by application
of a single, global, time-varying control.    Quantum computers
are devices that process information using quantum coherence and
entanglement.    They can build up arbitrarily complicated computations
by performing quantum logic operations on a few variables
at a time.   Techniques of electromagnetic resonance allow
one to construct quantum computers from arrays of coupled quantum
systems by applying a single, time-varying global control field [6-8].   
Here, we show that the same technique can be used to make the quantized
versions of integrable systems compute: the action-angle variables
of the integrable systems become the logical degrees of freedom of
the quantum computer, and a global control allows the systematic
coupling and manipulation of those variables to perform universal
quantum computation.

The dynamics of a classical Hamiltonian integrable
system can be decomposed into action-angle variables, so that the
motion of the system represents a trajectory on a set of tori in a
$2n$-dimensional phase space.
The KAM theorem shows that these trajectories are relatively stable
under perturbation [1].  
The action variables $I_j$ have vanishing Poisson brackets with the
Hamiltonian $H$ of the system: $\{ H, I_j\}=0$, and so are conserved.
Their Possion brackets with each other also vanish:
 $\{I_j, I_k\}=0$.   The angle variables $\theta_j$ rotate at frequency
$\omega_j = \partial H/\partial I_j$.  If the $\omega_j$ are incommensurate,
the state of the system covers an invariant torus in phase space.

In the quantized version of an integrable system [2-5],
the Hamiltonian and action-angle variables become operators and the Poisson
brackets are replaced by commutators.   Because they all commute
with each other, the Hamiltonian and the action variables can
be simultaneously diagonalized and have joint eigenstates
$|i\rangle = |E_i\rangle|i_1\rangle \ldots |i_n\rangle$.  
Suppose that our system is initialized in such an eigenstate 
$|\psi_0\rangle = |E^0\rangle|i^0_1\rangle \ldots |i^0_n\rangle$.
Look at the eigenspace ${\cal H}_0$ spanned by eigenstates with eigenvalues
that lie within a range $\Delta E$, $\Delta i_1$ $\ldots$
$\Delta i_n$ of the eigenvalues of $|\psi_0\rangle$.
Because the KAM theorem tells us that the classical dynamics are
periodic and robust under small perturbations,
we can approximate the quantum Hamiltonian over ${\cal H}_0$ by
$$H= H_0 + \sum_j{\partial H\over\partial I_j} \Delta I_j 
+ \sum_{jk} {\partial^2 H \over \partial I_j \partial I_k}
\Delta I_j \Delta I_k,\eqno(1)$$
where ${\partial H\over\partial I_j} = \omega_j$ as above,
and $\Delta I_j,\Delta I_k$ are harmonic oscillator Hamiltonians
restricted to ${\cal H}_0$.
That is, in the subspace spanned by eigenstates in the vicinity
of our starting state $|\psi_0\rangle$, the quantized version of the
integrable system behaves to first order like a collection of 
uncoupled harmonic oscillators, with couplings and nonlinear 
behavior that appear at second order. 

It is well-known how to make such quantum systems compute [6-8].
The quantized oscillators become the qudits of our quantum
computer.   By driving at the individual oscillator frequencies
$\omega_j$, and taking advantage of the second-order nonlinearity,
one can induce any desired transformation of the oscillators
individually.   By driving at the difference frequencies
$\omega_j-\omega_k$, one induces two-qudit transformations
between the $j$th and $k$th oscillators.   Taken together,
such one- and two-qudit operations allow one to perform
quantum computation.

Make this argument precise.
Add a global control field, with a time-varying strength: 
$$H(t) = H + \gamma(t) H_c.\eqno(2)$$
The state of the system $|\psi(t)\rangle$ evolves under the time-dependent
Schr\"odinger equation,
$i{\partial |\psi\rangle/ \partial t} = H(t) |\psi\rangle$.
First consider periodic driving, $\gamma(t) = \gamma_0 \cos\omega t$, 
and go to the interaction picture.  The state in the interaction
frame is $|\chi\rangle = e^{iHt} |\psi\rangle$,
which obeys the equation
$$i{\partial |\chi\rangle \over \partial t} = 
\gamma_0 \cos\omega t ~ e^{iHt} H_c e^{-iHt} |\chi\rangle.\eqno(3)$$
Write
$H_c = \sum_{i i'} a_{ ii'} 
| i\rangle \langle  i'|$
in terms of eigenstates $|i\rangle$ of $H$. 
In this basis, equation (3) becomes
$$i{\partial |\chi\rangle \over \partial t} = \sum_{ i i'}
(\gamma(t) a_{ i i'}/2)( e^{i(E_i - E_{i'} - \omega ) t}
+ e^{i(E_i - E_{i'} + \omega)t } | i\rangle\langle i'| ~ 
|\chi\rangle.\eqno(4)$$
We go to the rotating-wave approximation by dropping oscillating
terms, retaining only those where
$E_i - E_{i'}  \pm \omega \approx 0  $.
In the rotating-wave
approximation, $|\chi\rangle$ obeys the equation
$$i{\partial |\chi\rangle \over \partial t} = \tilde H_c |\chi\rangle,\eqno(5)$$
where $\tilde H_c = (1/2)\sum_{ii': E_i - E_{i'}  \pm \omega \approx 0}
\gamma_0 a_{ii'} |i\rangle\langle i'|$.   
The periodic driving then drives on-resonant transitions
with Rabi frequency $\gamma_0 a_{ii'} $ in the co-rotating frame.
In $SU(2)$ notation, if the two energy eigenstates
of the on-resonant transition are identified with 
spin-$z$ $\uparrow$, $\downarrow$, then we can drive 
rotations $e^{-i\gamma_0 t \sigma_{\hat\jmath}/2}$,
where $\hat\jmath$ is a vector in the $x-y$ plane in the
co-rotating frame whose phase is determined by the phase
of the sinusoidal driving term.  By varying this phase,
we can construct any desired $SU(2)$ transformation in 
the two-level subspace of the on-resonant transition.
Assume that the $\omega_j$ are incommensurate (for a 
typical chaotic system, their distribution is Poissonian [2-5]).
Varying the frequency of the driving field to drive 
different transtions affords full and systematic
control of the integrable quantum system. 

Now analyze the size of the errors introduced by the rotating-wave
approximation.  To zero'th order in perturbation theory, the energies of the
off-diagonal terms oscillate, yielding an average phase of zero.
To first order, the off-resonant transitions undergo small, rapid oscillations
with frequency
$ \Omega = \sqrt{ \gamma_0^2
+ \Delta \omega^2}$,  where $\Delta \omega^2 = (E_i - E_{i'}  \pm \omega)^2 $,
and amplitude $O(\gamma_0^2/\Omega^2)$. 
These are the terms ignored in the rotating-wave approximation:
they can be made as small as desired by driving with weaker and
weaker fields.    The effectiveness of the rotating-wave approximation
is justified by the KAM theorem [1]: the effect of the small perturbations
on the action-angle variables is simply to perturb
their frequencies.  For variables on resonance with the field, the
perturbation induces a controllable unitary evolution.  For off-resonant
variables, the oscillating perturbations average to zero.
Because we have assumed the frequencies of the action-angle variables
to be incommensurate, the length of the 
pulses required to perform selective driving to the desired
accuracy goes as $O(n)$: the time it takes to perform selective
driving grows linearly in the size of the system.

The method of selective driving allows one
to perform universal quantum control of 
integrable systems.  To perform universal quantum computation, 
we must be able to control the states of individual action-angle
variables and couple pairs of variables efficiently.
Begin in the ground state $|0\rangle \otimes \ldots \otimes
|0\rangle$, and drive transitions only in the ground
and first excited states $\ell=0, \ell=1$ for each oscillator.
These form the qubits of the universal computation.
The anharmonic spectrum of the Hamiltonian, together with the
assumption that their frequencies are incommensurate,
implies that all single qubit transitions have unique frequencies
and can be driven selectively using the global drive.
So frequency selection allows us to perform arbitrary
single qubit $SU(2)$ transformations on individual oscillators.
Driving at the frequency $\omega_j - \omega_k$, 
performs $SU(2)$ transformations on the $|10\rangle,
|01\rangle$ subspace of the $j$'th and $k$'th oscillator subsystem --
i.e., one can continuously `swap' the $j$'th and $k$'th
qubits.    (Alternatively, because the $I_j$ all commute
with the Hamiltonian, one can emulate NMR quantum information
processing [7-8] and use delays between resonant pulses
together with the interactions between action variables to
to effect $\Delta I_j\otimes \Delta I_k$ operations.)  Single-qubit rotations
and two-qubit continuous swap operations 
are universal for quantum computation. 

If the control Hamiltonian drives all transitions with
equal strength, so that $a_{ii'}$ is a constant, then we are
done: quantum integrable systems can be made to perform
universal quantum computation with a single global driving field.
In general, however, the transition
frequencies $\gamma_0 a_{ii'}/2$ depend on the states
of the other oscillators that are not involved in the single-oscillator
and two-oscillator transitions.  That is, the transition
driven at frequency $ \omega_j$ for the $j$'th
actually corresponds to a band of transition frequencies
oscillator
$\gamma_0 a_{ii'}$, where the $i,i'$ label not only the
state of the $j$'th oscillator, but the states of the other
oscillators as well.  To cope with this variation, make
the driving field of the form $\gamma(t) = 
\gamma_0(t) \cos \omega t$, where $\gamma_0(t)$ is a slowly
varying envelope field with bandwidth smaller than the frequency
difference $r\omega_j$ but large enough to cover the band of transition
frequencies $\gamma_0 a_{ii'}$.  
Adjust the strength of the component of
$\gamma_0(t)$ at each of these frequencies within the band to compensate for the
variation in the couplings $a_{ii'}$.  This adjustment can be performed,
for example, using optimal control techniques [9-10].   In [11] it is shown
that the complexity of the control problem for integrable systems scales
polynomially with the number of variables in the system.
As in decoupling pulses in NMR
quantum computing [7-8], the length of the compensating pulse 
grows with the number of action-angle variables.
In practice, the global
control may couple each action-angle variable to a
few others, so that only a small variation in transition frequencies
need be compensated for.  The same technique allows us to
compensate for variation in transition frequencies when coupling
two oscillators.

Combined with the ability to 
prepare and measure the state of at least one of the
oscillators, the ability to perform continuous time global control 
allows universal quantum logic via frequency selection.
The method is similar to quantum computation via electromagnetic
resonance [6-8].  As in NMR quantum computation [7-8], the global nature
of the control and coupling leads eventually to the `forest of lines'
problem: the number of spectral lines that one wishes to resolve grows
with the number of oscillators, so that the selective pulses must
become weaker, longer, and more exact in frequency as the number
of oscillators increases.  More precisely, as the number of action-angle
variables $n$ becomes large, the time it takes to perform each
selective pulse goes as $O(n)$, so that the time required to
perform $m$ operations scales as $O(mn)$ rather than as $n$.     
This polynomial slowdown can be resolved if the action-angle
variables correspond to variables that are spatially localized,
so that the now no-longer global control field can be 
applied to only a few oscillators at a time.   

\bigskip\noindent{\it Quantum computation and chaotic systems}

Comparing the control of quantized integrable systems with that of
quantized chaotic systems [2-5], we see that the method of resonant control
fails.  The spectrum of quantized chaotic systems typically obeys
a Wigner-Dyson distribution, so that the separation between 
energies in the spectrum decreases exponentially in the system size.
Although we can still use resonance methods to
drive transitions between energy eigenstates of the chaotic
system and to build up arbitrary unitary transformations, the
length of the selective pulses now grows exponentially with the
size of the system.  For example, we can impose a qubit structure on 
a chaotic quantum system by labeling the energy eigenstates
using binary numbers (e.g., label them sequentially from
smallest to largest). 
However, the spectrum will not obey the simple form of pairwise
coupled qubits or qudits as in equation (1).  Instead, the interactions
between qubits will be highly non-local: for a typical chaotic
system they obey the same statistics as a random Hamiltonian
[2-5].   In this case,
controlling qubits individually takes an exponentially large time. 
Although quantized
chaotic systems are generically controllable via a time-varying
global field [12], their controllability does not mean that quantized
chaotic systems can be made to perform universal quantum computation.

If the chaotic system is close to integrable,
in the sense that it possesses
a natural tensor product structure with weak interactions amongst
spectrally resolvable variables, then we can use resonant
quantum control selectively to decouple and recouple those
variables, leading to universal computation as above.  The chaotic
nature of the system will introduce noise into the system at a rate
given by the Kolmogorov-Sinai entropy: as long as this rate is
sufficiently small then bang-bang control or
quantum error correction techniques can
compensate for it in principle.  (Quantum error correction can also
compensate for noise induced by the interaction of an integrable
system with an environment.)   For a fully
chaotic, strongly coupled system, however, with no natural
tensor product structure, global control does not suffice
to perform quantum computation.

\bigskip\noindent{\it Discussion:}  Because integrable systems can
be decomposed in terms of action-angle variables, the quantized
version of such a system consists to first order of non-interacting quantized
harmonic oscillators, in which nonlinearities and couplings enter at
second order.  The theory of resonant driving then shows that
a global time-varying control field can be used 
selectively to drive individual oscillators and couple pairs
of oscillators to perform universal quantum computation.  By contrast,
quantum chaotic systems do not possess a natural tensor product
structure.  Although it may be controllable via a global
field, there is no obvious way to make a strongly
chaotic quantum system perform universal quantum computation.

\vfill

\noindent{\it Acknowledgements:} 
S.L. was supported by DARPA, AFOSR, ARO, and Jeffrey Epstein.
S.M. was supported by the DFG via SFB/TRR21 and by the EU projects 
SIQS and DIADEMS. 

\vfil\eject

\noindent{\it References:}

\smallskip\noindent
[1] V.I. Arnold, {\it Mathematical methods of classical mechanics,}
Springer, New York (1978).

\smallskip\noindent 
[2] F. Haake, {\it Quantum signatures of chaos,} 3rd edition, Springer, New York
(2010).

\smallskip\noindent
[3] Martin C. Gutzwiller, {\it Chaos in Classical and Quantum Mechanics,}
Springer, New York (1990).

\smallskip\noindent
[4] P. Cvitanovi\'c, R. Artuso, P. Dahlqvist, R. Mainieri,
{\it Chaos classic and quantum,} epublication: http://chaosbook.org/
  (2002).

\smallskip\noindent
[5] V. E. Korepin, N. M. Bogoliubov, A. G. Izergin, {\it Quantum Inverse 
Scattering Method and Correlation Functions,} Cambridge University Press,
Cambridge (1997).

\smallskip\noindent
[6] S. Lloyd, {\it Science} {\bf 17}, 1569-1571 (1993).

\smallskip\noindent
[7] D.G. Cory, A.F. Fahmy, T.F. Havel,
{\it Proc. Nat. Acad. Sci.} {\bf 94}, 1634-1639 (1999).

\smallskip\noindent 
[8] D.G. Cory, R. Laflamme, E. Knill, 
L. Viola, T.F. Havel, N. Boulant, G. Boutis,
E. Fortunato, S. Lloyd, R. Martinez, C. Negrevergne, M. Pravia, Y. Sharf, G.
Teklemariam, Y.S. Weinstein, W.H. Zurek, {\it Fort. der Physik} {\bf 48}, 
875-907 (2000).

\smallskip\noindent
[9] A.P. Peirce, M.A. Dahleh, H. Rabitz, {\it Phys. Rev. A} {\bf 37},
4950 (1988).

\smallskip\noindent
[10] P. Doria, T. Calarco, S. Montangero,
{\it Phys. Rev. Lett.} {\bf 106}, 190501 (2011). 

\smallskip\noindent
[11] S. Lloyd, S. Montangero, {\it Phys. Rev. Lett.} {\bf 113}, 010502 (2014).

\smallskip\noindent
[12] S. Lloyd, {\it Phys. Rev. A} {\bf 6202}, 2108 (2000).

\vfill\eject\end